# A measurement based software quality framework

Zádor Dániel Kelemen[1], Gábor Bényász[1,2] [*], Zoltán Badinka[1]

[1]ThyssenKrupp Presta Hungary Ltd.
1519 Budapest Pf. 531, Hungary
{Daniel.Kelemen, Gabor.Benyasz, Zoltan.Badinka}@ThyssenKrupp.com
[2]Corvinus University of Budapest
1093 Budapest, Fővám tér 8, Hungary

**Abstract**
In this report we propose a solution to problem of the dependency on the experience of the software project quality assurance personnel by providing a transparent, objective and measurement based quality framework. The framework helps the quality assurance experts making objective and comparable decisions in software projects by defining and assessing measurable quality goals and thresholds, directly relating these to an escalation mechanism. First results of applying the proposed measurement based software quality framework in a real life case study are also addressed in this report.

## 1    Introduction

Quality can be assured from multiple viewpoints and various processes can be put into focus. Debugging, testing, peer reviews, process audits and assessments, measurement and risk based quality improvement are common examples of assuring software quality.

In this report we focus on measurement, which can help strengthening the quality processes or products.

One common job of quality assurance experts to assess the quality the customer releases of a software project. In many cases they have the right to decide the go/no go of a customer release. This decision is especially important in automotive, medical, aerospace and safety critical projects where errors can cause loss of human lives. Without having a systematic approach in these scenarios the decisions made by quality assurance experts are based only on their experience and insight. Consequently, experience-based quality assurance may not provide the same results if it is performed by different personnel and/or in different projects. This is due that human decisions may differ depending on the background, situation and point of view on quality.

---

[*] Correspondence to gabor.benyasz@thyssenkrupp.com.

Moreover, even if the decision is supported by measurement, different metrics, different data collection and visualisation methods can affect and make the decision of quality assurance experts and the comparison of projects difficult.

In order to eliminate the problem of experience-only based quality assurance (and to unify the way of assuring quality among quality experts) we propose an objective and repeatable approach to quality assurance which is based on measurement, proposes unified metrics and goals through projects and decreases the dependency on the experience of the quality assurance experts. We call our proposed solution Measurement based Software Quality Assurance Framework (MSQF). MSQF consists of the (1) MSQF concept (MSQC) and (2) an MSQ Process (MSQP). We call MSQC the concept of repeatable software quality assurance driven by measurement with decreased dependency on the experience of the quality assurance experts and project settings. The application of MSQC helps achieving a transparent, organization-wide, objective measurement program based on shared and unified metrics, measurement data collection measurement and calculation. Such an approach can help organisations in achieving objective insight into the quality of processes and work products. In order to make the MSQC operational, we propose an MSQ process, which can help the implementation of the MSQC. The process itself provides a guidance and thus can be modified or tailored to organisational and project settings.

Besides increasing objectivity through software projects it is also a common need to comply with industry standards. Compliance to industry standards is both an internal goal and a requirement from customers. Thus, the definitive goal of MSQF is to provide a quality assurance approach which satisfies most relevant industry standards.

In section 2 this report is positioned in the literature. In section 3 the research approach is addressed. In section 4 a criteria for MSQF is identified. Section 5 provides an overview of the MSQP and section 6 provides details on the application of MSQF in an organisation and a project. Section 7 provides validation against MSQF criteria. Limitation and conclusion are presented in sections 8 and 9 respectively.

## 2 Background

A vast literature is available on software measurement and the state of the practice is evolving [1]. Many of the papers focus (1) on the process – the way measurement could be done. A number of publications focus on (2) how a measurement or process improvement program shall be introduced into an organisation and on the experiences of the introduction and (3) others focus on the meaning of metrics providing newer solutions as the technology evolves and data is available on the usage of metrics.

Some of the notable measurement concepts of the first set are the Goal/Question/Metric[2] and its extension the GQM+[3], the Experience Factory[4], the Practical Software Measurement[5] and the ISO/IEC 15939 [6] standard for measurement process among many others. Introducing a measurement program to an organisation (2) is also a topic of measurement-oriented literature [7]–[10]. (3) Discussions and debates on usefulness of and introduction of new metrics is a continuously evolving field [11]–[20].

Since we propose a concept (MSQC) with a process (MSQP) our solution can be positioned mainly in the first set, and it fills an important gap by emphasizing not just the process steps but the objectivity, organisational-wide transparency, repeatability, reduced dependency on the experience of the quality assurance personnel.

# 3 Research approach

Based on the introduction of MSQF in section 1, our research question can be formulated as follows: *Does an MSQF provide objective quality assurance through the project lifecycle and does it satisfy relevant industry standards?*

The research question can be divided into sub-questions. These questions will be answered by research steps each producing a research deliverable. Questions, steps and deliverables are summarized in Table 1.

Table 1 – Research questions, steps and deliverables.

| Research question | Research step | Deliverable | Section |
|---|---|---|---|
| Q1 What criteria shall an MSQF satisfy? | S1 Define criteria for MSQF | D1 MSQF Criteria | 4 |
| Q2 What process can be used in supporting the practical implementation of the MSQC? | S2 Design an MSQP | D2 An MSQP | 5 |
| Q3 How can we provide proof of concept for the MSQF? | S3.1 Validate the MSQF in a case study | D3.1 Case study summary on the application of the MSQF in a real-life project | 6 |
| | S3.2 Assess if the MSQF satisfies predefined criteria | D3.2 Assessment of the MSQF versus the MSQF criteria | 7 |

# 4 Criteria for an MSQF

Two major criteria were identified for (applying) MSQF: (1) provide objective quality assurance through the project lifecycle, (2) satisfy relevant industry standards. In this section we address these two criteria.

**(1) Provide objective quality assurance through the project lifecycle**

This criterion has been identified based on internal requirement and current experiences within projects. Project participants and project quality leaders were asked how they would improve the current quality assurance approach. Based on this input most of the respondents suggested that in most of the cases decision of the project quality leader relies only on his/her experiences and not on objective measurement data. Escalation was also undefined and there was a need to relate this to the measurement approach. Thus the objectivity criterion was refined and subdivided into criteria as follows:

C1. Provide objective quality assurance

    C1.1 Provide measurement based quality assurance,

    C1.2 Apply unified metrics in projects,

    C1.3 Set targets (thresholds) for metrics through project lifecycle,

    C1.4 Provide threshold-based escalation,

    C1.5 Share measurement results through the organisation.

    **(2) Satisfy relevant industry standards**

We considered the industry standards as mainly an external need. Customers usually assess software quality assurance systems based on two well-known and widely accepted quality approaches [21], [22]: SPICE [23] or CMMI [24].

Depending on the field, variants of SPICE may be important to organisations such as Automotive SPICE [25], Medical SPICE [26] or Enterprise SPICE [27] among others. Similarly, in case of CMMI constellations of the model such as CMMI for Acquisition [28] or CMMI for Services [29] may be considered important.

We performed a quick comparison of these industry standards into their variants and we came to the conclusion that no significant difference exists among them from measurement point of view.

Since we will perform a case study at an automotive company, most relevant criteria are:

    C2.1 Satisfy Automotive SPICE MAN.6 Process area,

    C2.2 Satisfy CMMI-DEV Measurement and Analysis Process Area.

## 5    An MSQ process

After defining the MSQF criteria (C1.1-C1.5, C2.1-C2.2) in section 4, we define a possible MSQP which can help making the MSQC operational. Due to quality control purposes, the MSQP is strongly related to the escalation mechanism. The MSQ presented here can be extended or tailored to organisational and project needs.

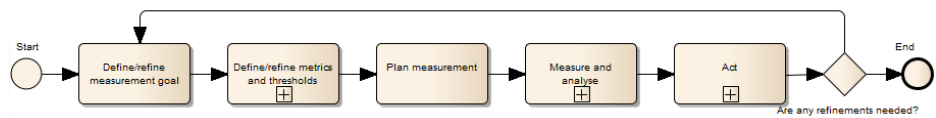

Figure 1 – An MSQ process

The MSQP consists of the following activities/sub-processes:

**1.    Define/refine measurement goal**

Measurement goals (objectives) shall be defined in accordance with the organizational/business goals and measurement information needs.

**2.    Define/refine metrics and thresholds** (see further details on Figure 2)

This sub-process includes the following activities: Define metric, Define data visualization, Define abstract milestones, Define measurement data collection and Define thresholds and escalation.

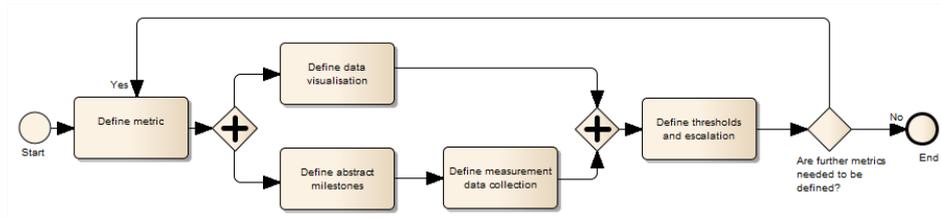

Figure 2 – Define Metric

## 3. Plan measurement

Planning measurement involves both the organisational and project level planning of measurement activities. This can affect the quality measurement strategy, the quality metrics and the project level quality goals.

## 4. Measure and analyse (see further details on Figure 3)

This sub-process includes the following activities: assuring the availability of measurement data, collecting measurement data and the analysis of measurement data.

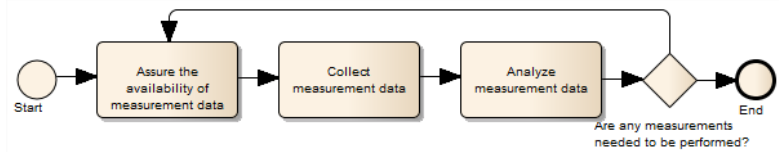

Figure 3 – Measure and Analyze

## 5. Act (see further details on Figure 4)

This sub-process focuses on the communication/sharing the measurement results, in case of deviations identification of root causes, escalating violations of thresholds and taking alternative actions such as improving the measurement approach, refining metrics and thresholds.

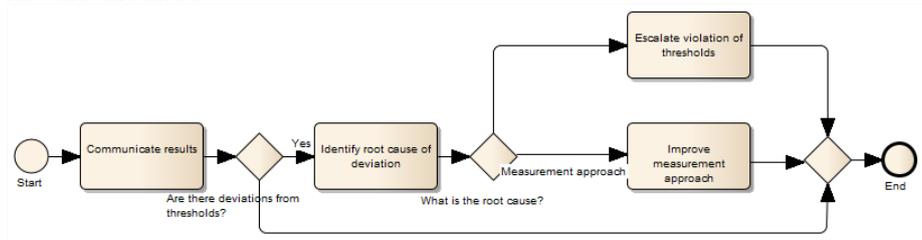

Figure 4 – Act

***Notions related to the MSQP:***
**Escalation mechanism:** The goal of the quality escalation strategy is the handling of deviations observed during measurement in order to enable early correction.

**Abstract milestones:** By an abstract milestone we mean a checkpoint in a project where the quality of a project is needed to be assessed. Abstract milestones provide the basis for measurement data sampling and escalation. Abstract milestones are generally applicable to all of the metrics and are dependent on organizational and project settings. Abstract milestones can overlap major (customer) releases.

**Deviations** from the quality plan can easily be defined in percentage from the goal (threshold). The higher the deviation the higher the risk of occurrence of quality problems.

**Escalation levels at abstract milestones:** Quality measurement based escalation can be defined on multiple levels by connecting them to the deviations from the thresholds (e.g. depending on the percentage of deviation). The number of escalation levels and their relation to the thresholds can be defined according to the organizational needs. For examples of escalation levels, thresholds and abstract milestones see Table 2, Table 4 and Figure 5 in the appendix.

**Improvement of thresholds:** Thresholds shall be revisited on regular basis and updated based on organizational changes and needs.

# 6    Application of the MSQF in a real-world scenario

Various work products, roles and activities may be present at organisations. Thus, a tailoring of the MSQP may be needed to organisational settings. We decided to tailor the theoretical MSQF to organisational needs of ThyssenKrupp Presta Hungary (TKPH).

**About the company**

TKPH's website provides a good overview on the company: "The ThyssenKrupp Presta Group is a technology leader in the field of steering systems and a major innovative partner of the auto industry.

We are working at 16 sites around the world on developing modern technology solutions that make vehicle transport safer for all of us. The Budapest affiliate of the Liechtenstein-based ThyssenKrupp Presta company has been dealing with the development of the electronic steering systems of passenger cars since 1999 as the only electronic and software development competence centre within the company group. We are committed to satisfying demands of the largest auto manufacturers in the world with innovative, high-quality products. Our goal is to develop a safe and efficient steering system that provides a perfect driving experience to our clients."[30] TKPH has both internal and customer software projects, all of them focusing on developing electronic steering systems and has about ~300 employees at Budapest site.

Due to the limitations of this report we do not discuss all the details of the application of MSQF, we rather focus on the most important tailoring and deviations in implementation. Details not addressed here were performed as described in MSQF.

We discuss the tailoring of the MSQP to organisational settings (6.1), applying it in a project (6.2), then deviations in implementation on organisational (6.3) and project level (6.4).

### 6.1 Tailoring of MSQF to TKPH organisational settings

**Definition of metrics**

At the beginning of the measurement process definition, a goal was targeted that at least one metric shall be defined for each Automotive SPICE HIS scope process. In order to perform this, brainstorming sessions were held with the relevant stakeholders (i.e. process owners and process performers in projects). Metrics and their parameters such as metric definition, way of data collection, metric responsible were defined and prioritized. At the end of metric definition period a consensus was reached on the selected metrics to be measured. Table 5 shows an examples of project metrics defined on organizational level (to be applied in all projects) categorized by Automotive SPICE HIS scope process areas.

**Organisation specific work products**

In order to implement the MSQP in practice we had to take into consideration, update or define organisation specific work products. Most important related work products were:

MSQP tailored to TKPH – the measurement based quality assurance process has been tailored to TKPH settings extending the basic process with TKPH specific work products and adding additional processes in order to practically sustain the measurements.

Quality assurance strategy – includes strategic level descriptions of quality goals, measurement approach and escalation mechanism.

Project quality plan – Each project has a quality plan which focuses on the tailoring of organisational processes to project needs and quality goals/thresholds based on metrics.

Release audit report – Each customer release is subjected to a release audit, where the project quality leader and other project participants judge the quality and safety of the release. The release audit report is the output of the release audit.

Measurement database – A unified measurement tool and database has been introduced at TKPH.

Quality metrics – This document contains the software quality metrics to be applied at TKPH.

MSQF training – a training material has been developed and used for employees on the about measurement program and the measurement tool

**Organisation specific roles**

The following roles are applied at TKPH: Project Quality Leader (PQL) – plans, tracks and controls the quality of the processes and products in customer projects, independent from the project organisation. Project Leader – leads the customer projects, Quality Department Leader – line manager, leads the quality department, including the work of PQLs.

**Escalation levels at abstract milestones**

The escalation mechanism is defined on three levels: 0-2, which are depending on abstract milestones and the percentage of deviation from thresholds. Table 2 summarizes the escalation levels by the deviation at the beginning of the project and the targeted role of escalation. A relatively high deviation is allowed at the beginning of the project, while deviations at the end of the project are not allowed, thus escalation levels are monotonically decreasing from 30/20/10% to 0%.

Table 2 – Escalation levels and deviations from the thresholds

| Escalation level | Deviation at project start | Deviation at project end | Escalated from | Escalated to |
|---|---|---|---|---|
| Level 0 | ±10% | 0% | Project Quality Leader | Project Leader |
| Level 1 | ±20% | 0% | Project Quality Leader | Quality Department Leader |
| Level 2 | ±30% | 0% | Quality Department Leader | Higher Level Management |

**Tailoring of the 'Improve measurement approach' activity:** if the 50% of the projects fit within the escalation level 2 thresholds (the widest, with ±30-0% deviation) and more than 25% of the projects fit within the escalation level 0 (in the range of ±10-0% deviation) then the threshold will not be changed. In other cases actions are needed to be taken (e.g. modification of the thresholds or communication of the importance of the metric or redefinition of the metric).

**Unified metrics, Data collection, data integration**

For the unified data collection and data integration we used DataDrill Express from Distributive Management [31]. This tool has two main components (1) a data collector (2) a data visualization. With the collector we collected data from various systems such as IBM Rational DOORS, IBM Rational Change, Excel and CSV files and various types of databases through ODBC among many others. Prerequisites of common metrics, data collection and data calculation were that all projects use the same tools following the same processes and an agreement was reached on the common set of metrics and way of data calculation.

## 6.2    Applying MSQF in a project

In this section we provide a brief introduction of the basic project settings and how the measurement program was introduced.

Due to confidentiality reasons we cannot share the project name (thus we call it Project A) nor concrete measurement results (therefore the measurement data is only for demonstration purposes).

The goal of the project A is to develop reusable software modules to the automotive customer projects. The introduction of the MSQF into project A started after a year of the project starting date, at Release 2 (mid 2013). Previously only release candidates and nightly builds were available and releases were not subjected to release audits. The project was working in an agile way in 3-week sprints which later changed to 2-weeks sprints in order to be aligned with other projects of the organization. The project was planned to be finished at Release 4.0.0. The average time between major releases in this project is three months. These releases are subjected to release audits.

The MSQF was introduced together with basic quality assurances work products of the company (including the application of the quality strategy, project quality plan and release audits). This additional work products were discussed in the organisation level tailoring and help easing the implementation in TKPH settings.

Within the introduction of MSQF, metrics and related thresholds were identified and introduced, all related to ASPICE HIS scope process areas, including requirements engineering, through problem and change management to test management among others. Well-known metrics such as requirements coverage, test coverage, open change requests or open defects were identified. For source code quality a subset of HIS metrics [32] were selected. Measurement data collections were performed on a weekly basis and analysed by the project quality leader.

Table 4 in appendix shows a possible example of quality thresholds and escalation levels related to different milestones of a project for *open defects* - a metric used related to Automotive SPICE Problem Resolution Management process area. The goal in this case was to reduce the number of defect to zero by the end of the project.

Figure 5 in Appendix shows an example of measurement data visualization in a project. A similar visualisation is used in Project A. On Figure 5 escalation levels across the project lifecycle are represented by lines: red (escalation level 2), orange (escalation level 2) and yellow (escalation level 0) and threshold/goal is represented by the green line. The rest of the figure shows the open defects in various stages (verifying, implemented, assigned etc.) within the project in a weekly breakdown from calendar week 34/2013 to 26/2014, textually indicating only every 6th week for better visualization.

### 6.3 Deviations in implementation on organisational level

A major problem we encountered in applying MSQF in multiple projects was resulted by the limited abilities of the tool we used. After piloting the MSQF in Project A, we included a number of internal and customer projects (10+) into the measurement program. We encountered practical problems both in data collection and data visualization. In data collection the collector interface did not work correctly when a high amount of data was needed to be collected, therefore it was rewritten for collecting data in multiple cases e.g. collecting from IBM Rational Change or IBM Rational DOORS. From visualization point of view the number of reports (figures) was planned to be generated on a weekly basis exploded (10 projects x 15+ metrics). In this part of the tool we encountered serious difficulties of parameterization and thus the number of weekly generated figures grew over 300.

All of the deviations or issues encountered on implementing MSQF on organizational level were practical which can be solved by alternative (BI) tools.

### 6.4 Deviations in implementation on project level

One major deviation from MSQF happened within the project A. Namely, it was difficult to set goals/thresholds for the metrics which were closely related to the software modules (e.g. review coverage of the requirements, or branch and MC/DC coverage).

These metrics were measurable on module level and directly reflected the quality and completeness of the module. Therefore during release planning at setting the new quality and project thresholds for the next release, high level goals were broken down to the module level and weighted by module size, and module quality improvements were planned according to their priority. Priority was defined by external projects which were requiring to reuse the modules developed by project A. The breakdown of the high level quality goals to module level was proposed and performed by the software project leader. This was an important practical addition to the proposed MSQF, and strengthened the feasibility and way the project prioritization and the quality goals has been aligned release by release.

## 7    Assessment against MSQF criteria

Table 3 provides a brief assessment of MSQF in the case study against the criteria identified in section 4.

Table 3 – Assessment of the MSQF against MSQF criteria

| Criterion | Assessment of compliance to criterion |
|---|---|
| C1. Provide objective quality assurance | |
| C1.1 Provide measurement based quality assurance | The basic concept of MSQF is measurement. In order to make it operational, the MSQP defines most important measurement activities (e.g. metric definition, data collection, analysis etc.) For more details see Figure 1, Figure 2, Figure 3 and Figure 4. As it is defined so it was applied in the case study. |
| C1.2 Apply unified metrics in projects | In the definition of the MSQC included that unified metrics shall be used in the project to make them comparable and transparent. In the case presented metrics, thresholds, way of data collection, data calculation and analysis were defined on organizational level and also related to escalation mechanism. Despite that both the concept and the case study included the notion of unified metrics, and same metrics were applied in project A as in other projects, in order to strengthen the practical implementation of MSQF, it may be useful to add this aspect to the MSQP as a separate activity. |
| C1.3 Set targets (thresholds) for metrics through project lifecycle | The MSQP includes an activity "Define thresholds and escalation" (see Figure 2) covering C1.3. In the case study we performed this activity first on organizational level by defining thresholds and escalation mechanism related to abstract milestones. In Project A, we tailored the thresholds and abstract milestones to project settings by aligning thresholds to the major releases of the project and by setting the thresholds taking into the account (1) the situation at release planning, (2) desired result for the next release and (3) and targeted results at the final milestone of the project. |

| Criterion | Assessment of compliance to criterion |
|---|---|
| C1.4 Provide threshold-based escalation | MSQP includes an activity for escalation "Escalate violation of thresholds" (see Figure 4). Furthermore, the MSQF also includes the investigation of the root cause of the deviation and taking possible alternative actions such as improving the measurement approach.<br>In Project A major deviations from the planned goals happened in 3 cases. 2 out of 3 deviations were escalated according to the escalation strategy. By analysing the root cause of the third deviation, the project came to the conclusion that a major re-planning is needed due to the lack of resources. This was reasonable and accepted by higher level management and it resulted in postponing the end-date of the project. Consequently, the new project milestones were supported by quality with new thresholds and goals. |
| C1.5 Share measurement results through the organisation | MSQP includes an activity for sharing measurement results called "Communicate results" (see Figure 4).<br>In the organisation measurement results are shared through the measurement tool on a weekly basis, Project A shares its results through the tool to which the relevant stakeholders have access. |
| C2. Satisfy the measurement-related criteria of industry standards | |
| C2.1 Satisfy Automotive SPICE MAN.6 Process area | The theoretical MSQP does not fully satisfy all requirements of the ASPICE MAN.6 nor CMMI-DEV MA, because these quality approaches include requirements which are not measurement specific but to be applied to all processes (see e.g. generic practices). These generic requirements will always depend on the organisational process structure, therefore will not be included into the MSQP. |
| C2.2 Satisfy CMMI-DEV Measurement and Analysis Process Area | The MSQP tailored to TKPH location satisfies all specific and generic practices of CMMI Measurement and Analysis process area as well as all base and generic practices of ASPICE MAN.6 For a high level mapping see Table 6 in the Appendix. |

# 8 Validity and reliability of the research

In this section the focus is on validity and reliability of the research: addressing reliability, construct, internal and external validity.

Since the MSQF is applied in a real world case study, we do not address ecological validity [33], [34].

We also exclude internal validity. According to Yin, "internal validity [is] (for explanatory or casual studies only and not for descriptive or exploratory studies): seeking to establish a causal relationship, whereby certain conditions are believed to lead to other conditions, as distinguished from spurious relationships" [34]. In this study we did not investigate if certain conditions led to other conditions, thus according to Yin's definition internal validity is not relevant in our research.

**Construct validity**

Strategies for ensuring construct validity are (1) multiple sources of evidence (also called triangulation), (2) thick descriptions (also called chain of evidences) and (3) review of the draft case study report by key informants [33]–[35].

In order to assure construct validity we relied on the second option: thick description of the chain of evidences. In the case study the whole process is documented starting from the development of MSQF through the weekly measurement results till escalation. Additionally, release planning and release audit reports thoroughly document the results and the conclusions the quality expert and the project made at certain milestone. Unfortunately these cannot be shared publicly, but they are shared within the organisation.

Despite of these documentation we acknowledge that the case could have been described in more details without the restrictions of the confidentiality settings. Agreeing on release status by the project and contributing to the development of the MSQF by breaking down high level goals to module goals also provided a clear option for reviewing results of implementing the MSQF, this is slightly related to the third tactic proposed by Yin, the review of the results by key informants.

**External validity**

External validity is "generalisations or interpretations that a researcher has proved in a particular context apply equally well to other populations of other contexts" [36]. Since we propose a practical solution and performed a case study in a single organisation and a single project it is important to address external validity.

The main issue here is to prove that the MSQF is applicable at other organisations or projects. In order to assure this, we defined the MSQ concept and the MSQ process at the highest possible abstraction level, moreover we discussed how we tailored the concept and the process and related work products to organisational and project settings.

We must acknowledge that applying MSQF at other organisations will require strong management support, openness and a good measurement data collection and visualisation tool support. For applying the MSQF in other projects we already have preliminary results (not covered within the frame of the case study presented in this report), namely: we have already added the full set of metrics to two customer projects and we are introducing them to two further customer and an internal project. Therefore we can confirm that if it is accepted on organisational level it is workable in multiple different projects.

**Reliability**

Yin identifies two tactics for reliability: (1) usage of a case study protocol "to deal with the documentation problem" and (2) the development of a case study database. He also adds: "a good guideline for doing case studies is … to conduct the research so that an auditor could in principle repeat the procedures and arrive at the same results" [34].

In order to ensure repeatability we documented the case from 4 viewpoints: (1) documentation of the MQSP (2) documentation of the results (on organisation level:

TKPH-tailored measurement process, quality strategy, quality metrics document, on project level: quality planning, tailored thresholds  and release audit reports), (3) documentation of the decisions taken in release audit reports and (4) – the case settings, tailoring and most important deviations and issues on organisational and project level (in section 6). Despite that we tried to document the case from 4 interrelated viewpoints, it could be documented in a more detailed report which would include more detailed description of steps, communication aspects and a summary of concrete release thresholds and results, however this report provides insights for applying MSQC and also provides a possible MSQP which due to its high abstraction level can be followed/tailored by other personnel achieving the same results. Abstracted examples in the appendix also support the better understanding the practical repeatability of the MSQF.

## 9    Conclusion and further steps

In this report we presented a practically usable solution to the objective and transparent quality assurance which decreases the dependency on the experience-only based quality assurance. This approach is called the Measurement Based Software Quality Framework (MSQF). The MSQF consists of the concept (MSQC) and a process (MSQP). The concept was validated in a real organisation and a real project. A possible process is provided to support the practical implementation of the concept which can modified or tailored to organisational and project settings based on needs and goals. We have also defined a criteria for MSQF which was based on our internal needs (C1) – mainly focusing on the reduced dependency on the experience of the QA staff, and external requirements – coming from relevant industry standards (C2). Beside the case study, we also discussed the compliance of the MSQF to the MSQF criteria. Here a major deviation was that the proposed MSQP itself does not satisfy all requirements of the industry standards, but the tailored version was thoroughly mapped and it satisfies both standards identified. Finally, we addressed the reliability and validity issues of this research.

Further steps of this research will include refinements of the MSQF based on the first case study and C1/C2 and also the implementation of the MSQF in multiple projects of the organisation and possibly also at other organisation.

# 11    Appendices

## 11.1    Thresholds and escalation levels example

Table 4 – Thresholds and escalation levels related to open defects in the pilot project

| Mile-stone name | Thresh-old | Esc level 0 (%) | Esc level 0 | Esc level 1 (%) | Esc level 1 | Esc level 2 (%) | Esc lev-el 2 |
|---|---|---|---|---|---|---|---|
| Release 2.0.0 | 500 | 10 | 550 | 20 | 600 | 30 | 650 |
| Release 2.1.0 | 400 | 8 | 432 | 16 | 464 | 24 | 496 |
| Release 2.2.0 | 300 | 6 | 318 | 12 | 336 | 18 | 354 |
| Release 3.0.0 | 200 | 4 | 208 | 8 | 216 | 12 | 224 |
| Release 3.1.0 | 100 | 2 | 102 | 4 | 104 | 6 | 106 |
| Release 4.0.0 | 0 | 0 | 0 | 0 | 0 | 0 | 0 |

## 11.2    Example of visualizing measurement data, thresholds and escalation levels

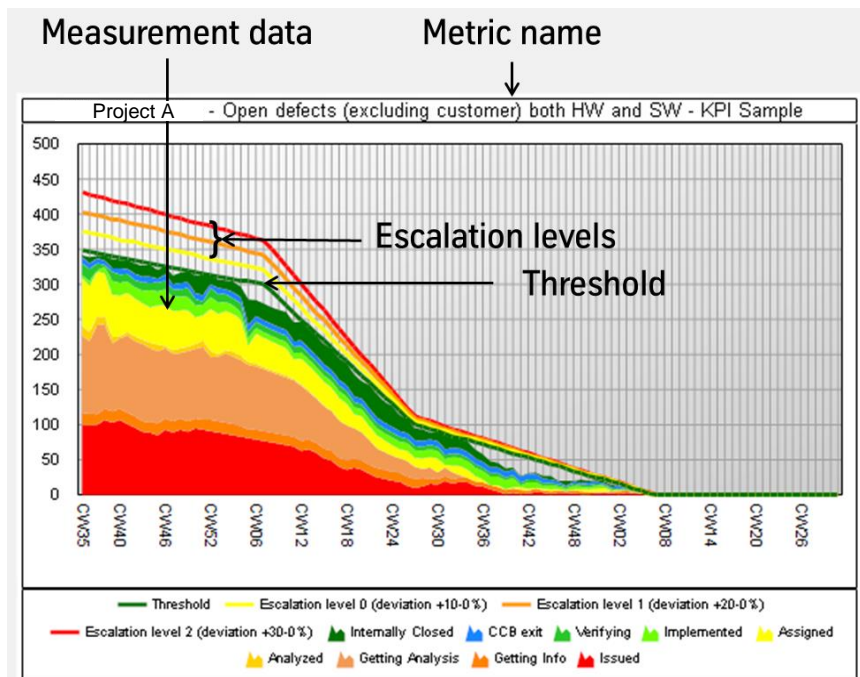

Figure 5 – Demonstration of visualizing measurement data, escalation levels and thresholds tracked on a weekly basis

## 11.3 Examples of metric definitions

Table 5 – Examples of metrics defined on organizational level

| Metric name | Definition (all metrics are measured within projects) |
|---|---|
| **SUP.9 Problem Resolution Management** | |
| Open defects (excluding customer) both HW and SW | The metric shows the planned versus actual number of open internal (excluding customer) hardware and software defects in a weekly break-down with the following statuses: (1) issued, (2) analyzing, getting analy-sis, assigned, implemented, verifying, getting info, internally closed. The metric is a state snapshot including all issues being in one of the states mentioned above. |
| Open customer defects | The metric shows the planned versus actual number of open hardware and software customer defects in a weekly breakdown with the following statuses: (1) issued, (2) analyzing, getting analysis, assigned, implement-ed, verifying, getting info, internally closed. The metric is a state snap-shot including all issues being in one of the states mentioned above. |
| **SUP.10 Change request management** | |
| Open change requests (exclud-ing customer - internal) | The metric shows the open internal (excluding customer) change requests planned for the current or for the previous customer release with the following statuses: (1) issued, (2) analyzing, (3) exit CCB, getting analy-sis, assigned, implemented, verifying, getting info, internally closed. |
| Open customer change requests | The metric shows the open customer change requests which: (a) are ordered by the customer or promised to the customer, (b) and are planned for the current or for the previous customer release with the following statuses: (1) issued, (2) analyzing, (3) exit CCB, get-ting analysis, assigned, implemented, verifying, getting info, internally closed. |

## 11.4 MSQP tailored to TKPH settings

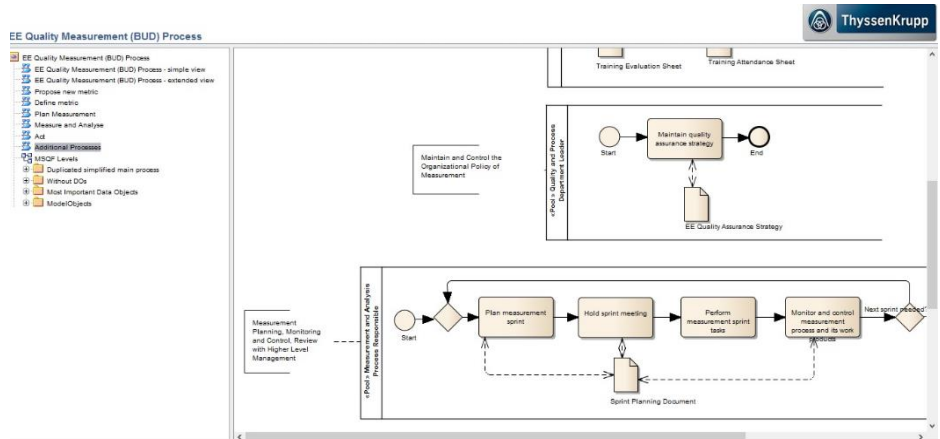

Figure 6 – Screenshot of the HTML export of the TKPH measurement BPMN process model created in Enterprise Architect

## 11.5   Process element mapping to CMMI and ASPICE

Table 6 – MSQF process elements tailored to TKPH settings and mapped to CMMI and ASPICE

| Element name | Stereotype | CMMI-DEV MA | ASPICE MAN.6 |
|---|---|---|---|
| EE Quality Assurance Strategy | DataObject | MA GP 2.1 | MAN.6.BP1 MAN.6.BP2 |
| Maintain quality assurance strategy | Activity | MA GP 2.1 MA GP 2.6 | |
| Hold sprint meeting | Activity | MA GP 2.10 MA GP 2.3 MA GP 2.4 | MAN.6 GP 2.1.4 MAN.6 GP 2.1.5 MAN.6 GP 2.1.6 MAN.6 GP 2.2.1 MAN.6 GP 2.2.2 MAN.6 GP 2.2.3 |
| Communicate results | Activity | MA GP 2.10 MA.SP 2.4 | MAN.6.BP9 |
| Plan measurement sprint | Activity | MA GP 2.2 | MAN.6 GP 2.1.1 MAN.6 GP 2.1.2 MAN.6 GP 2.1.5 MAN.6 GP 2.2.3 |
| Sprint Planning Document | DataObject | MA GP 2.2 MA GP 2.6 | |
| Quality Department (not represented on figures in Appendix) | Pool | MA GP 2.4 | |
| Train employee | Activity | MA GP 2.5 | |
| Measurement and Analysis Training Material | DataObject | MA GP 2.5 | |
| Training Attendance Sheet | DataObject | MA GP 2.5 | |
| Training Evaluation Sheet | DataObject | MA GP 2.5 | |
| Collect measurement data | Activity | MA GP 2.7 MA.SP 2.1 MA.SP 2.3 | MAN.6.BP6 |
| Monitor and control measurement process and its work products | Activity | MA GP 2.8 | MAN.6 GP 2.1.3 MAN.6 GP 2.2.4 |
| Evaluate adherence | Activity | MA GP 2.9 | |
| Start | StartEvent | MA GP 3.1 | |
| Propose new metric | Activity | MA GP 3.2 | MAN.6 GP 2.2.1 MAN.6.BP3 |
| Define/refine measurement goal | Activity | MA GP 3.2 MA.SP 1.1 | ENG.2-10.GP 2.1.1 MAN.3.GP 2.1.1 MAN.6.BP3 SUP.1.GP 2.1.1 SUP.10.GP 2.1.1 SUP.8.GP 2.1.1 SUP.9.GP 2.1.1 |
| Define/refine metrics and thresholds | Activity | MA GP 3.2 MA.SP 1.2 | MAN.6.BP4 |
| Define metric | Activity | MA.SP 1.2 | |

| Element name | Stereotype | CMMI-DEV MA | ASPICE MAN.6 |
|---|---|---|---|
| Define abstract milestones | Activity | MA.SP 1.2 MA.SP 1.3 | |
| Define measurement data collection | Activity | MA.SP 1.3 | |
| Measure and analyse | Activity | MA.SP 1.4 | MAN.6.BP5 |
| Analyze measurement data | Activity | MA.SP 2.2 MA.SP 2.3 | MAN.6.BP7 |
| Measurement and Analysis Database | DataStore | MA.SP 2.3 OPD.SP 1.4 | MAN.6.BP6 |
| Measurement and Analysis Process Responsible | Pool | | MAN.6 GP 2.1.4 |
| Improve measurement approach | Activity | | MAN.6.BP10 MAN.6.BP11 |
| Act | Activity | | MAN.6.BP8 |
| Take appropriate action | Activity | | MAN.6.BP8 |
| Project Quality Plan | DataObject | | SUP.1.BP1 |
| Escalate violation of thresholds | Activity | | SUP.1.BP10 |